\documentclass[twocolumn,article,superscriptaddress,groupedaddress]{revtex4}
\usepackage[export]{adjustbox}
\usepackage{booktabs}
\usepackage{units}
\usepackage{graphicx}
\usepackage{enumitem}
\usepackage{xcolor}
\usepackage{hyperref}
\usepackage{psfrag}
\usepackage{graphicx}
\usepackage{dcolumn}
\usepackage{bm}
\usepackage{textcomp}
\begin{document}


\title{First full cool down of the SPIRAL 2 superconducting LINAC}

\author{Adnan Ghribi}
\email{ghribi@ganil.fr}
\affiliation{Grand Acc\'el\'erateur National d'Ions Lourds (GANIL)}%
\affiliation{Centre National de la Recherche Scientifique (CNRS - IN2P3)}%
\author{Muhammad Aburas}%
\affiliation{Grand Acc\'el\'erateur National d'Ions Lourds (GANIL)}%
\affiliation{Centre National de la Recherche Scientifique (CNRS - IN2P3)}%
\author{Yoann Baumont}%
\affiliation{Grand Acc\'el\'erateur National d'Ions Lourds (GANIL)}%
\affiliation{Commissariat à l'Energie Atomique (CEA - IRFU)}%
\author{Pierre-Emmanuel Bernaudin}%
\affiliation{Grand Acc\'el\'erateur National d'Ions Lourds (GANIL)}%
\affiliation{Commissariat à l'Energie Atomique (CEA - IRFU)}%
\author{Stéphane Bonneau}%
\affiliation{Grand Acc\'el\'erateur National d'Ions Lourds (GANIL)}%
\affiliation{Centre National de la Recherche Scientifique (CNRS - IN2P3)}%
\author{Guillaume Duteil}%
\author{Robin Ferdinand}%
\affiliation{Grand Acc\'el\'erateur National d'Ions Lourds (GANIL)}%
\affiliation{Commissariat à l'Energie Atomique (CEA - IRFU)}%
\author{Michel Lechartier}%
\author{Jean-François Leyge}%
\author{Guillaume Lescalié}%
\affiliation{Grand Acc\'el\'erateur National d'Ions Lourds (GANIL)}%
\affiliation{Commissariat à l'Energie Atomique (CEA - IRFU)}%
\author{Yann Thivel}%
\affiliation{Grand Acc\'el\'erateur National d'Ions Lourds (GANIL)}%
\affiliation{Commissariat à l'Energie Atomique (CEA - IRFU)}%
\author{Arnaud Trudel}%
\affiliation{Grand Acc\'el\'erateur National d'Ions Lourds (GANIL)}%
\affiliation{Commissariat à l'Energie Atomique (CEA - IRFU)}%
\author{Laurent Valentin}%
\affiliation{Grand Acc\'el\'erateur National d'Ions Lourds (GANIL)}%
\affiliation{Centre National de la Recherche Scientifique (CNRS - IN2P3)}%
\author{Adrien Vassal}%
\affiliation{Grand Acc\'el\'erateur National d'Ions Lourds (GANIL)}
\affiliation{Commissariat à l'Energie Atomique (CEA/INAC-SBT)}%
\affiliation{Univ. Grenoble Alpes}%
\affiliation{Univ. Caen Basse Normandie}

\date{\today}

\begin{abstract}
SPIRAL 2 is a high intensity heavy ions beams accelerator project that has been going on for more than 10 years now. Countless efforts in different disciplines made it what it is today. One of the most important steps after the set up of the different equipments has been the very first full cool down of the superconducting cavities in an accelerator operation type configuration. While this has been a major achievement for the SPIRAL 2 teams, it also hi-lighted new challenges and constraints that would have to be addressed in order to have a high availability rate of the beam from the cryogenics side. This paper retraces this particular episode.
\end{abstract}

\pacs{Valid PACS appear here}
\maketitle


\section{\label{sec1}Introduction}
The GANIL's (Grand Acc\'el\'erateur National d'Ions Lourds) SPIRAL2 heavy ions accelerator\cite{Gales:2011he, Lewitowicz:2006fx,Bertrand:2007tp,Petit:2011ub,Ferdinand:2010ty} aims at delivering some of the highest intensities of rare isotope beams on earth. The unstable beams will be produced by the ISOL(Isotope Separation On-Line) method via a converter or by direct irradiation of fissile material. The driver is designed to accelerate protons, deuterons and heavy ions at different ranges of intensities and energies. It is composed of high performance ECR sources (Electron Cyclotron Resonance sources), a RFQ(Radio-Frequency Quadrupole), and a superconducting light/heavy ion LINAC(LINear Accelerator). The latter\cite{Junquera:2006vh,Ferdinand:2008ui,Bernaudin:2013tx,Ferdinand:2013ty} is based on 26 bulk Niobium made superconducting, independently phased resonators. These resonators are assembled in two families of cryomodules that integrate passive RF, vacuum and cryogenics components needed for their operation. The first family, called type A cryomodules and developed by CEA/IRFU, is optimised for $\beta_{0}=0.07$  relative velocity and comprises 12 cryomodules (one cavity/cryomodules). The second family, called type B cryomodules and developed by CNRS/IN2P3/IPNO, is optimised for $\beta_{0}=0.12$ and comprises 7 cryomodules (two cavities per cryomodule). All cryomodules are interspaced with room temperature quadrupole magnets and are operated at $\unit[88.0525]{MHz}$.

\begin{figure*}
  \includegraphics[width=15cm]{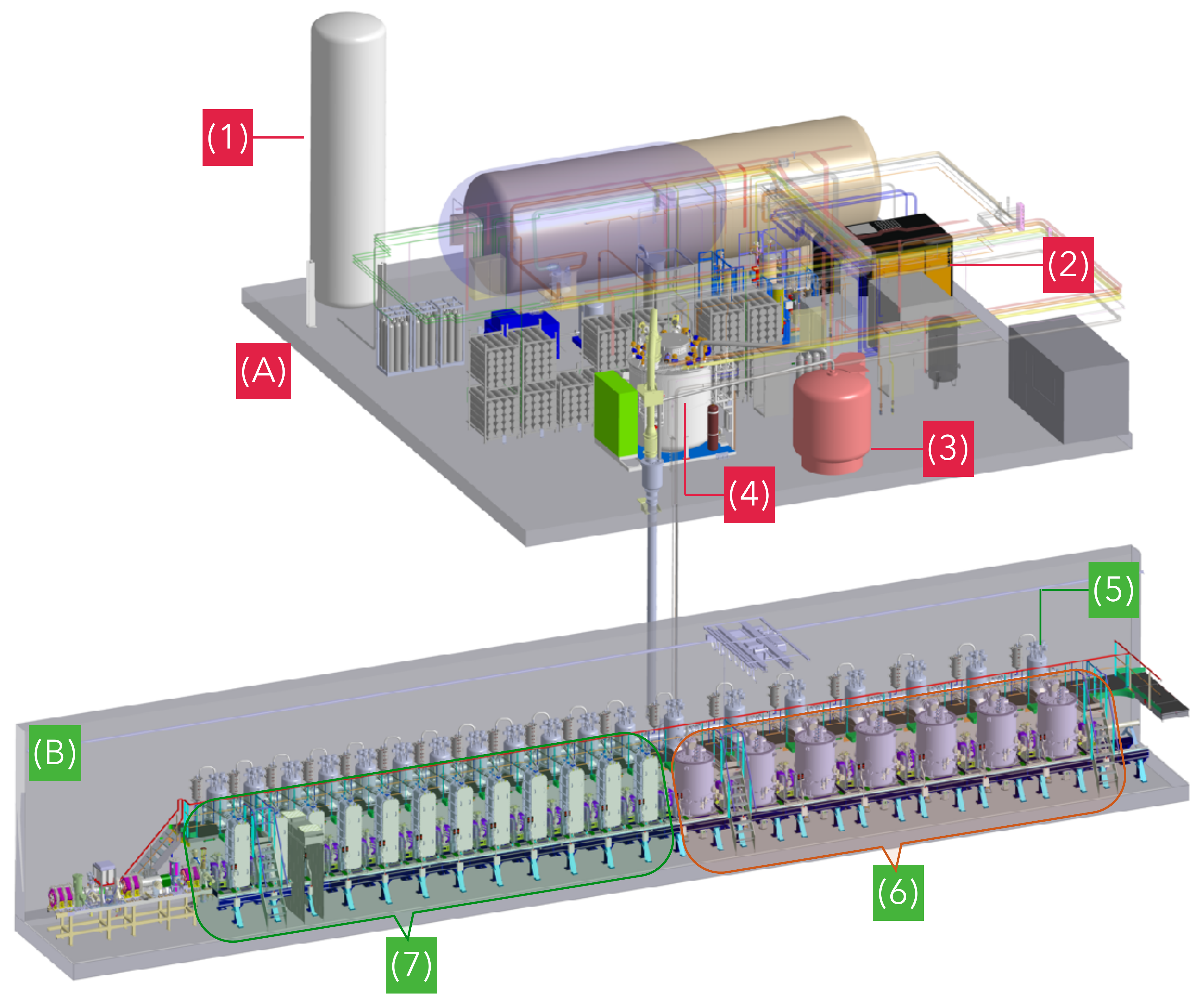}
  \caption{3D view of the cryogenic facility of Spiral 2. (A) Cryoplant. (B) LINAC. (1) LN2 tank. (2) Compressor station. (3) LHe dewar. (4) Cold box. (5) Valves-boxes. (6) Type B double cavities cryomodules. (7) Type A single cavity cryomodules.}\label{3d_sp2}
\end{figure*}

To be operated, the superconducting cavities need to be maintained at a stable temperature lower than their transition temperature with a stringent mechanical stability to avoid detuning. This is achieved thanks to a cryoplant delivering up to $\unit[1300]{W}$ cooling power at $\unit[4]{K}$. The coldbox, at the center of the cryoplant, delivers liquid helium through a phase separator (5000L capacity LHe dewar) and a cryodistribution system to the cryomodules and manages the return gas. For every cryomodule, a valves-box allows maintaining stable LHe and stable pressure at the surface of the cavities. An overall view of the cryogenic facility is visible on figure \ref{3d_sp2}. More details on the cryoplant can be found in \cite{GHRIBI201744}.

More than 10 years after the design phase have been necessary to build and assemble all the systems and subsystems. 37 km of cables link hundreds of sensors and actuators to 37 Siemens S300/S400 programmable logic controllers. Human Machine Interface uses PCVue\texttrademark~ for the cryoplant and WinCC\texttrademark~ for the LINAC. EPICS\texttrademark~ is used for data acquisition in redundant distant servers. In 2015, pre-commissioning of the coldbox confirmed its capability to deliver the required refrigeration power. In 2016, a first partial cool down of two cryomodules allowed the pre-commissioning of mechanical, piping, vacuum and control/command systems. A year later, a second partial cool down of three cryomodules allowed the commissioning of the command, control and acquisition system\cite{ghribi:in2p3-01569768}. The same year, all the building blocks including cryoplant and LINAC have been assembled with success.

The present paper brings the development and commissioning of the cryogenic system of SPIRAL 2 a step forward with a first full cool down of the LINAC. This is also certainly a step further into maturing the system for reliable beam operation conditions. In the first section, challenges and requirements for cool down and normal operation conditions are highlighted. The second section details the cryogenic process logic that has been implemented for the cool down. Next, we show the results of the cool down phase and the normal operation phase. An opening into RF measurements of the cavities properties is also described with a possible link to cryogenics.

\section{Cooling down the LINAC}

\subsection{Requirements and applied methods}\label{cool-down requirements}

\subsubsection{Global cool down strategy}
Due to the design of the cryoplant and the cryodistribution, a specific global cool down strategy has to be followed. As the cold box delivers the cooling power to cool down and maintain stable both thermal screens circuits (60 K) and liquid helium circuits, the cold box and the thermal screens are cooled down at the same time. Cool down of liquid helium circuits starts when the cold box reaches stable operation conditions and the main liquid helium dewar (5000 L capacity) has at least 1500 L of liquid helium. As there is no bypass at the edges of the LINAC, the edge cryomodules have to be cooled down in order to cool down the cryodistribution system. Once the cryodistribution is cold, the rest of the cryomodules follows in a specific order. A summary of the global cool down procedure used for 2017 cool down is shown below :

\begin{enumerate}
\item Purge of all systems
\item Pump the isolation vacuum and the beam vacuum
\item Helium liquefaction in the main dewar to a minimum of 1500 L.
\item Warm up the cold box
\item Cool down the cold box and at the thermal screens of CMA01 and CMB07 at the same time (the two edges of the LINAC).
\item  Once the cold box is in normal operation mode, cool down the cavities of CMA01 and CMB07. This concludes the cool down of the cryo distribution system. since there is no by pass at the ends of the LINAC, cooling down the two cryomodules at the end of the LINAC is mandatory to finalise the cool down of the cryo distribution system.
\item Cool down cryomodules from the edges to the center beginning by the thermal screens then the cavities for every cryomodule. Since the heat load of one type B cryomodule is approximately twice the heat load of a type A cryomodule, for every type B cryomodule, two type A cryomodule have been cooled down at the same time.
\end{enumerate}

 \subsubsection{Local cool down strategy}
  For every cryomodule, thermal screens are cooled down before the cavities. 2 valves ensure the cooling and control of the thermal screens and 3 valves ensure the cooling and control of the superconducting cavities. One of the three cavity valves (largest flow coefficient) is used in the beginning to cool down in a bottom to top way in order to take advantage of the cold vapours. Once, the cavity is cold and the local phase separator has a given level of liquid helium, the cool down valve is closed and the liquid helium level is regulated from the top of the phase separator using another lower flow coefficient valve. Details of the instrumentation of the cryomodules can be found in \cite{GHRIBI201744}. It should be noted that there are two outlet valves. The first is a warm one and is located at the outlet of a passive exchanger. When He gas is too warm ( > 80 K) to be re-sent to the cold box, it sends the gas back to the inlet of the compressor station. When the gas is sufficiently cold, the warm outlet valve is closed and a second one is opened. The latter is the one that usually controls the pressure of the cavities liquid helium baths. However, during all the phases of the cool down and until the cryomodules are cold and fully thermalised, this valves is kept entirely open. It should also be noted that the passage between the two mentioned valves can be sometimes tricky depending on the programmed speed of the cool down and the heat load of the cryomodules. This means that, when not programmed carefully or during trials, the passage can lead to security interlocks activation due to overpressure. Later on, a second level security has been added with a weight added on the inlet valves opening that depends on the cavities pressure and that overtakes the programmed cool down slope when the pressure reaches a certain value and before the activation of the security interlocks.
  
  It should be noted that, during cool down, the outlet valve that controls the pressure of the cavities liquid helium baths in normal operation modes is kept 100\% open at all times. 

One of the constraints of SPIRAL 2 cavities is the Q disease. This so called Q disease is a symptom of degradation of the quality factor of a cavity due to Hydrogen trapping on the inner surface\cite{Knobloch2003}. This appears if a cavity spends more than a given time $\Delta(t)$ between two critical temperatures T1 and T2. SPIRAL 2 cavities are not heat treated and suffer from such effects. During test bench qualifications, these values have been found to be : $\Delta (t)$ = 1 hour ; T1 = 50 K and T2 = 150 K. Fortunately, the symptom is reversible as hydrogen is freed during warm up when the temperature of a cavity increases. This means that if, for some reason, a cavity spends more than 1 hour in the critical temperature range, it has to be warmed up and cooled down again. Therefore, a specific cool down procedure has been used to avoid this situation. The strategy here is to slow down the cavities temperature drop in order to allow more time for thermalisation of the mass. Previous individual cool downs allowed to determine the temperature transmitter that represents the best the thermal behaviour of the cavity. This temperature has then been set as an observer for automatically controlling the slope of the temperature drop thanks to an active proportional integral (PI) controller. Between 300 K and 250 K, the main cool down valve is opened slowly to 100\%. Starting from 250 K, the PI controller on the same valve is activated with a sliding target cavity temperature from 250 K to 150 K. The slope of this variation is chosen by the operator.

\subsubsection{Results}

\begin{figure}[hbt]
  \includegraphics[width=.5\textwidth, right]{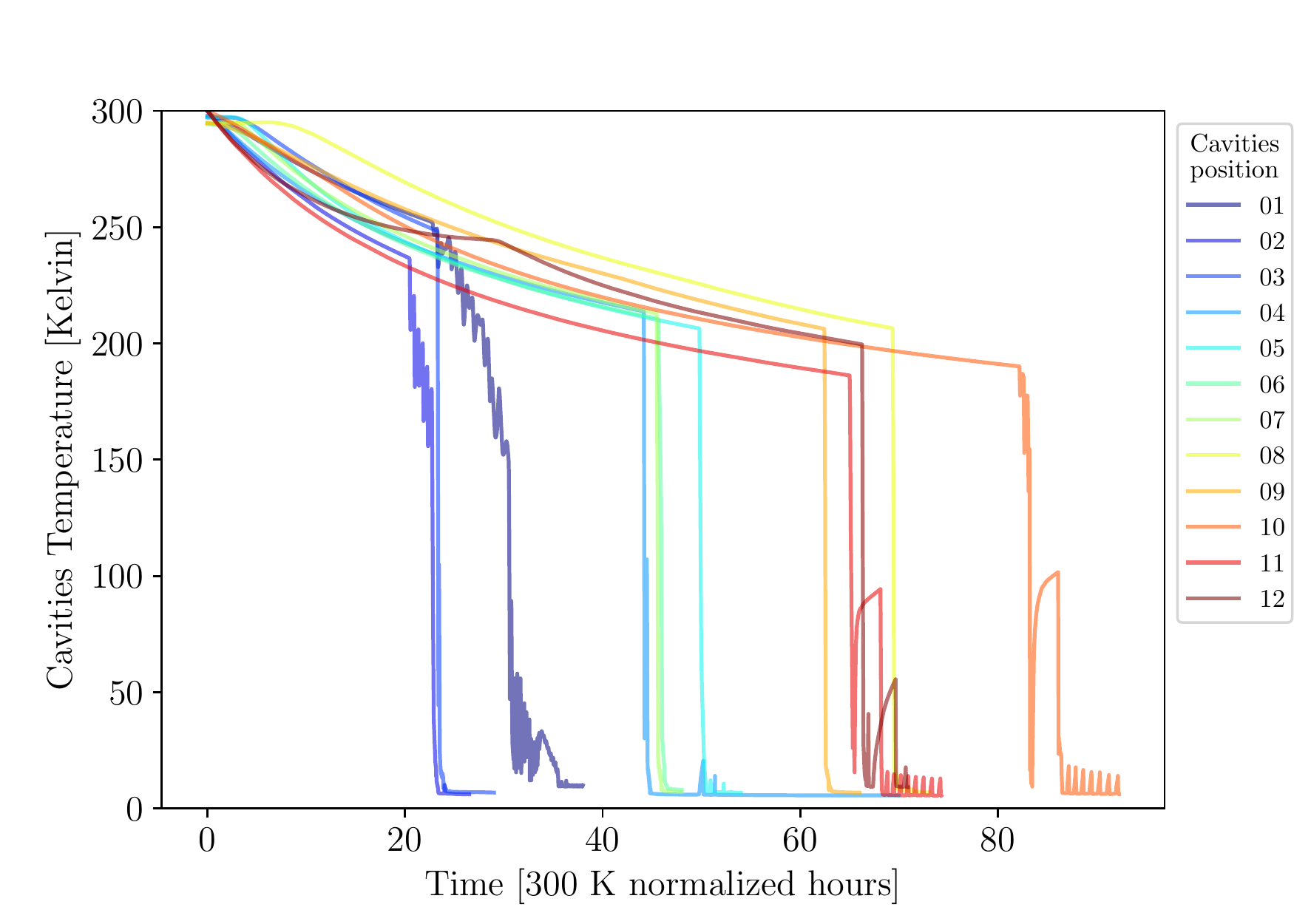}
  \caption{CMA cavities temperature drop. The time axis is normalised so that all temperature appear begin their slope at the same time }\label{cma_tcav_mef}
\end{figure}


\begin{figure}[hbt]
  \includegraphics[width=.5\textwidth, right]{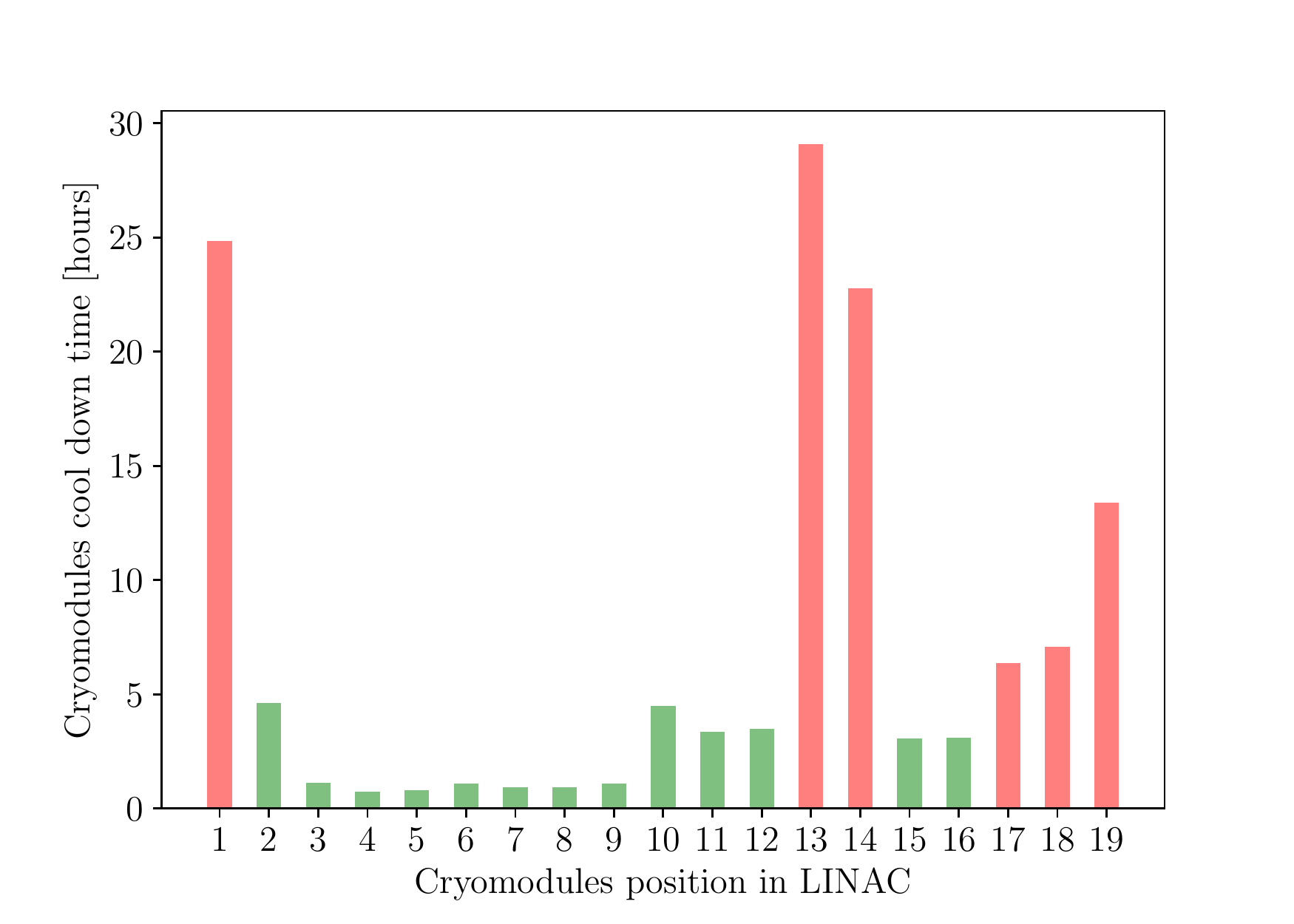}
  \caption{Cryomodules cool down time. Green bars are for cool down time lower than 5 hours. Red bars are for cool down time higher than 5 hours.}\label{CM_cooldown}
\end{figure}

\begin{figure}[hbt]
  \includegraphics[width=.5\textwidth, right]{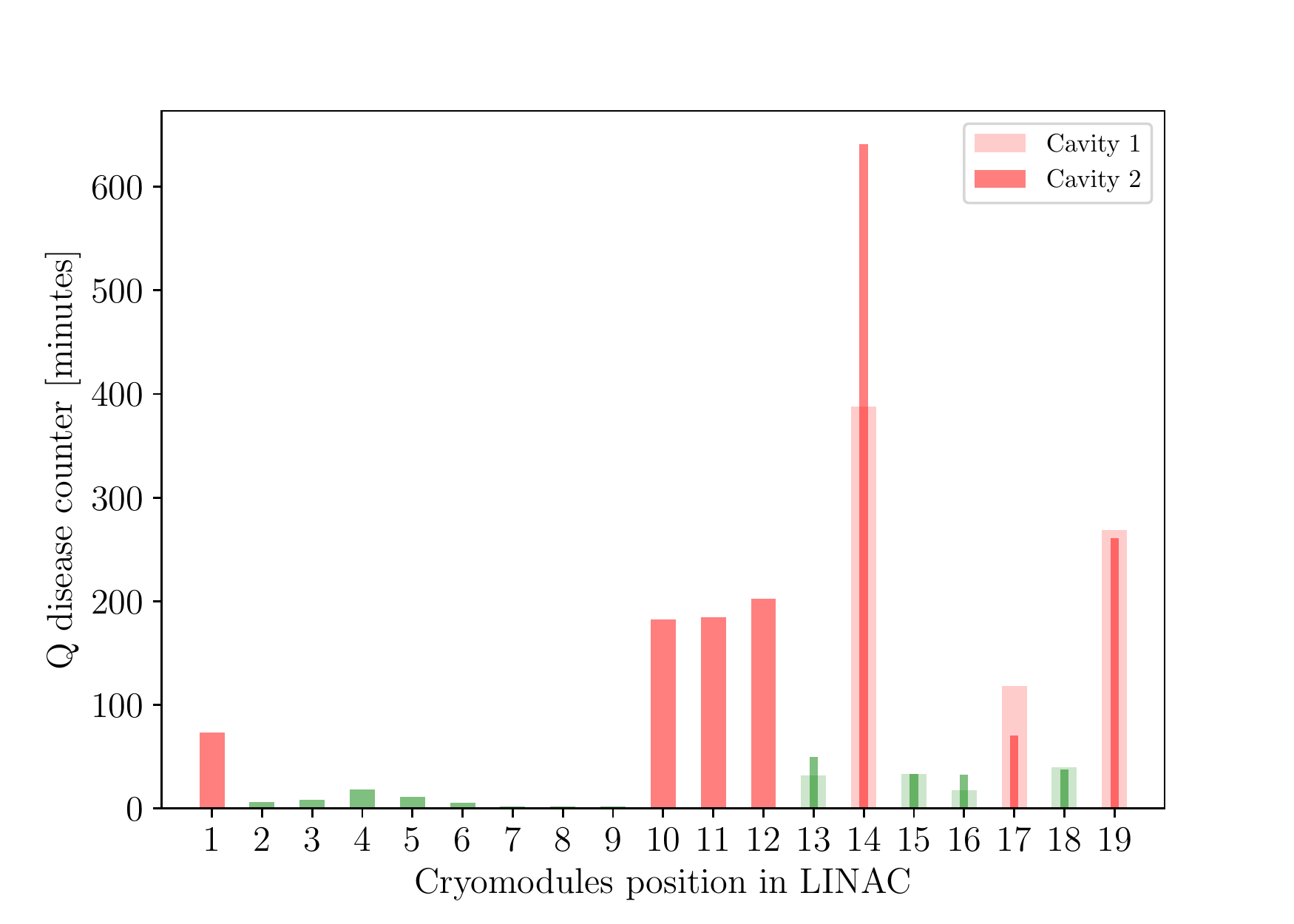}
  \caption{Q disease counter for all cryomodules. Green bars correspond to counters within the requirements. Red bars are for counters above requirements.}\label{qdcount}
\end{figure}

The procedure described in \ref{cool-down requirements} has been applied to oct. 2017 SPIRAL 2 LINAC cool down. It was the first full cool down of the LINAC and a good opportunity to experiment cool down methods and strategies. The resulting over all cool-down time is shown on figure \ref{CM_cooldown}. It begins at the start of the user controlled cavities cool down procedure and ends when the first percent of liquid helium is measured in the phase separators. As expected, the first cavities cool down time (cavities 1, 25 and 26) are higher because it matches the cool down of the liquid helium circuits of the cryodistribution. Cavities 13 and 14 (same cryomodule type B) are right in the middle of the LINAC, at the junction between the two main branches of the cryodistribution system. These cavities suffered from unequal control of thermal load distribution during cool down and not adapted cool down parameters. As a result, this deprived them from an adapted cold helium flow in a critical step of their cool down. Cavities 2, 10, 11 and 12 suffered from security mode passage. This is visible on the extended cool down time with respect to other type A cavities (the first 12 cavities) in figure \ref{CM_cooldown} as well as in the slope of their temperatures during cool down in figure \ref{cma_tcav_mef}.

As described previously cavities temperatures have been monitored in order to optimise cavities temperature slopes parameters. The temperature drops of the first 12 cryomodules can be seen in figures \ref{cma_tcav_mef}. In this figure, one can distinguish four main behaviours. The first is a curve visible on all cavities. It starts at 300 K and corresponds to a radiative cool down that occurs thanks to thermal screen cool downs. The second one is a change in the slope of the previous curve. It corresponds to the beginning of the cavity cool down procedure. The third is the step-controlled cavities temperature drop down. Its slope depends on the chosen slope parameters. In figure \ref{cma_tcav_mef}, the latter is visible only for cavities 1 and 2. The fourth is an acceleration of the cool down, again visible in figure \ref{cma_tcav_mef} for cavities 1 and 2. For most other cavities, the different parts can not be distinguished due to different reasons (beginning of cool down too fast, controlled cavities cool down too fast). For some cryomodules, the parameters entered were not adapted to the intrinsic heat load of the cryostats. This caused an increase in the outlet pressure, which caused in its turn a reduction of the cold helium flow and therefore an increase and stagnation of the cavities temperature in the critical temperature range.

The success or failure of a given cool down procedure or a parameter set is summarised in figure \ref{qdcount}. It represents an automatic counter $Qd_{count}$ that starts when the cavities temperature reach 150 K and stops when it is bellow 50 K. For cryomodules 1 and 7 (cavities 1, 25 and 26), that have been cooled down at the same time as the cryodistribution, the cool down has been slowed down because of the very high thermal load of the cryodistribution compared to the cryomodules. In addition, when the temperature of the helium return is lower than 80 K, it is redirected to the cold box cold return line. For the first cryomodules, the overall return temperature is too high for the higher cold stage of the cold box. Valves at the input of the cold box, therefore, significantly reduce the return flow while the temperature is too high. The other cases for cryomodules 10, 11, 12 and 14 (cavities 15 and 16) have been explained in the previous paragraph by accidental stop of the cool down. As for cryomodule 17 (cavities 21 and 22), it suffered from an input temperature slope too high for its thermal load.

\section{The way to operation}

\subsection{Thermalisation}
When the liquid helium level in the phase separators of the cavities is stable and regulated around a 90\%, the cool down procedure is finished. However, we have to wait until all systems and subsystems are thermalised before authorising cryogenic operation in the beam configuration. To know when this can happen, we regularly measured the heat load for every cryomodule by measuring the liquid helium level decrease during a given time when outlet valve is open and inlet valve is closed. The heat load is then given by :

\begin{equation}
Q = \frac{dV}{dt}\rho C_{l}
\end{equation}
where Q is the thermal load, V the LHe volume, t the time, $\rho$ the liquid helium density at the cryomodule measured pressure and $C_l$ the latent heat of liquid helium. For increased precision, the exact 3D volume of the phase separator have been modelled.

\begin{figure}[hbt]
  \includegraphics[width=.5\textwidth, right]{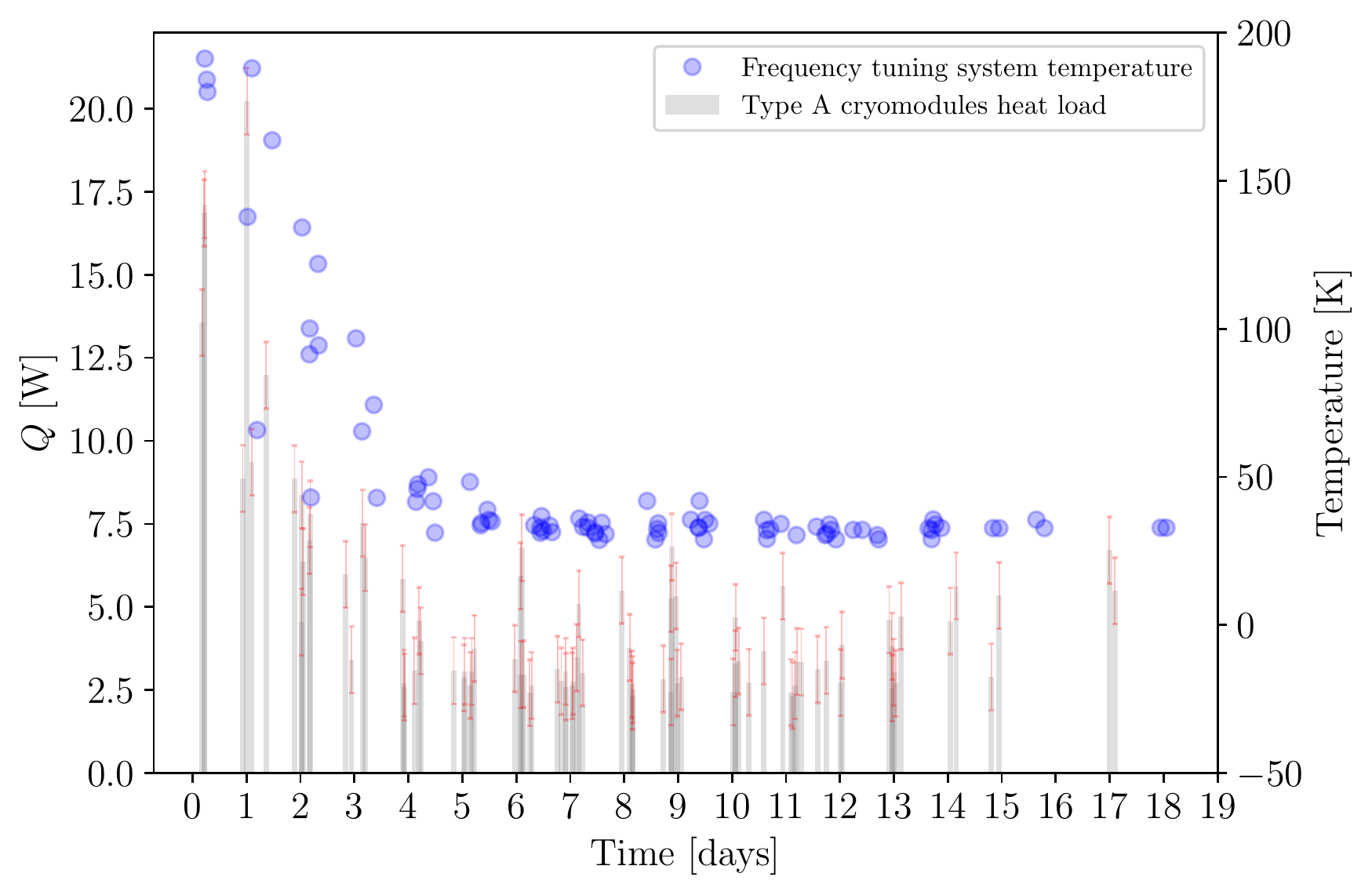}
  \caption{Bars : evolution of the measured thermal load $Q$ of the type A cryomodules (first 12 cavities) as a function of the elapsed time since the end of the cool down procedure.  Circles : Evolution of the temperature of the frequency tuning system.}\label{thermalisation_cma}
\end{figure}

\begin{figure}[hbt]
  \includegraphics[width=.5\textwidth, right]{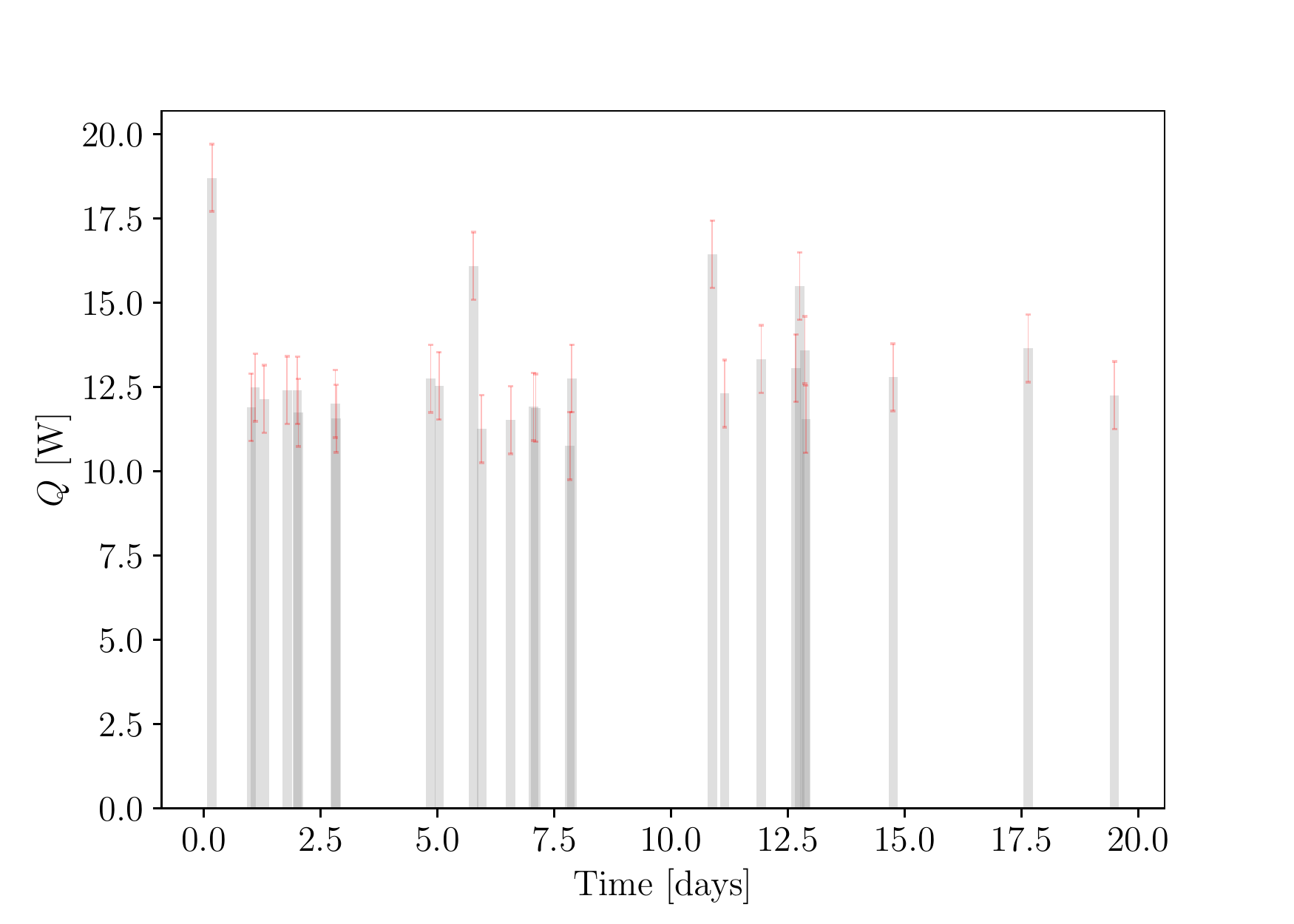}
  \caption{Evolution of the measured thermal load $Q$ of the type B cryomodules (last 14 cavities) as a function of the elapsed time since the end of the cool down procedure.}\label{thermalisation_cmb}
\end{figure}

What can be noticed from figures \ref{thermalisation_cma} and \ref{thermalisation_cmb} is that the time that we are looking for strongly depends on the considered cryostat geometry, construction and constraints. For example, type A cryomodules are fast to cool down but slow to thermalise (4 to 5 days) due to the presence of a specific frequency tuning system thermally decoupled from the cold mass (see figure \ref{thermalisation_cma}). The cryomodules are then thermalised when their frequency tuning systems temperatures reach a stable plateau (around 35 K). The type B cryomodules (see figure \ref{thermalisation_cmb}) have a much higher thermal load but are much faster to thermalise (around 1 day).

\subsection{Cavities liquid helium bath pressures and levels control}
Once the cryomodules are thermalised, there are two main constraints to authorise RF and beam operation. 

\begin{figure}[hbt]
  \includegraphics[width=.5\textwidth, right]{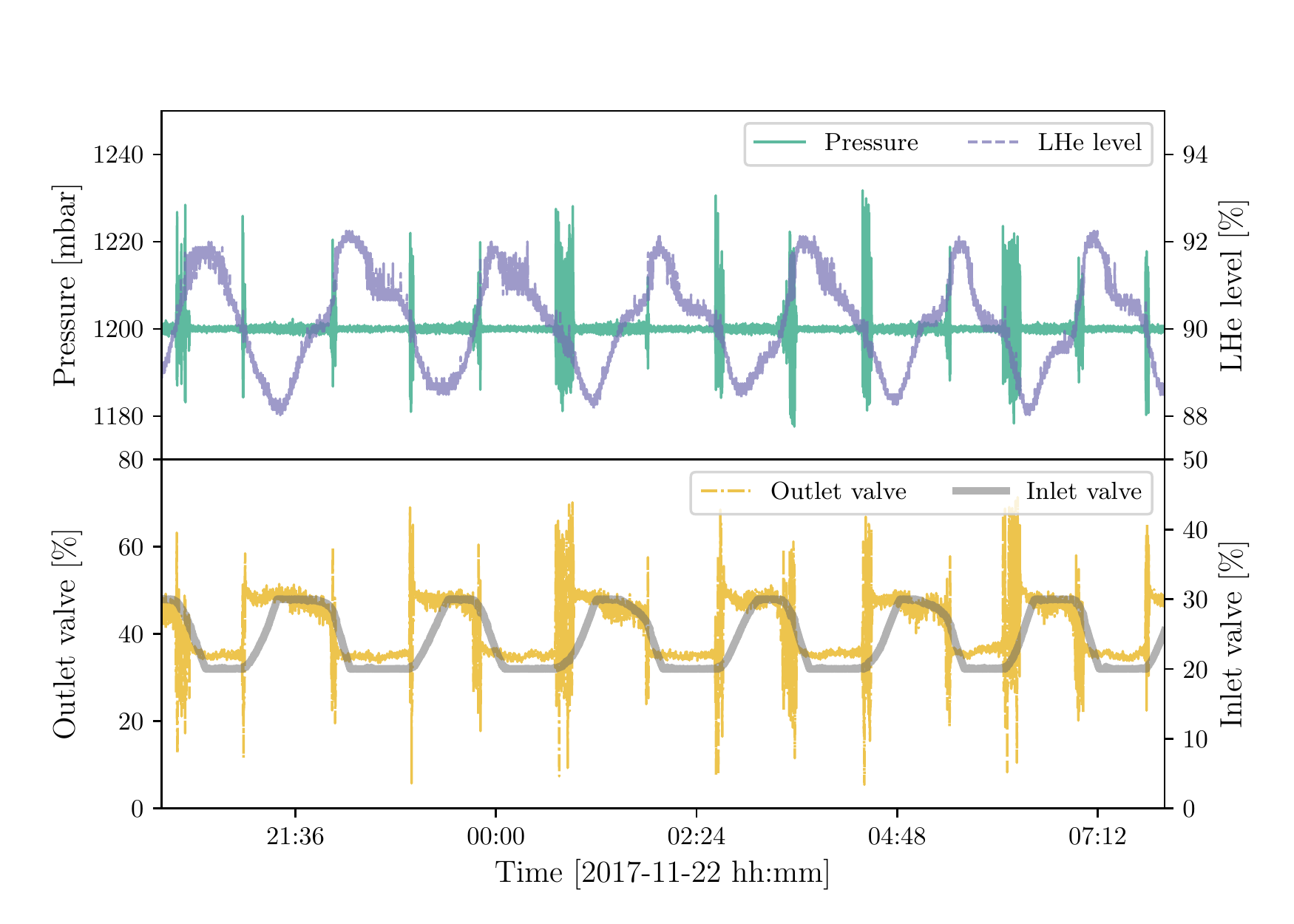}
  \caption{Pressure and liquid helium level regulations (top plot) with matching valves (bottom plot) for CMA05. Data of 2017-11-20.}\label{pt_lt_cma05}
\end{figure}

The first is to make sure that all cavities are submerged by liquid helium at all times. When the LHe bath pressure is stable, all cavities have a very stable temperature and therefore very stable RF properties. In order, to achieve this target, a dedicated valve is used to control liquid helium level. This valve uses a proportional-integral (PI) controller with an integral tuned to match the corresponding cavity static heat load and a proportional term chosen to make the valve slow enough in order not to have an instantaneous effect on the pressure. The pressure and liquid helium level of a same volume are obviously coupled. The matching valves that use independent PI controllers for liquid helium and pressure regulations therefore introduce correlations. As can be seen on figure \ref{pt_lt_cma05}, when not tuned properly, these correlations can be critical with regard to the stringent pressure stability requirements. The difficulty is enhanced here by the asymetric behaviour of the liquid helium regulation (fast to fill and slow to empty). One of the actions that has been done while keeping a simple PI controller was to limit the range of action of the liquid helium flow control valve. This proved efficient as it stabilised the pressure of the cavities with regard to correlated instabilities with liquid helium control valves positions. However, this approach is not adapted to a high variability of the thermal load.

The second is to make sure that pressure variations of the liquid helium bath do not exceed a given $\Delta(P)$. In fact, the forces that liquid helium apply on the surface of a cavity slightly changes its shape, which changes its geometric factor. As a result, its unloaded quality factor (that we can denote $Q_{fo}$ here) also changes as can be seen in this equation :
\begin{equation}
Q_{fo} = \frac{G}{R_s}
\end{equation}
where $G$ is the geometric factor and $R_s$ is the surface resistance.
Depending on the bandwidth of the cavity and the margin on the amplifiers side, this can be critical as it drastically reduces the capability to accelerate particles with the required acceleration gradients. For SPIRAL 2, the limit of $\Delta(P)$ has been set to $\pm~5~mbar$.

\begin{figure}[hbt]
  \includegraphics[width=.5\textwidth, right]{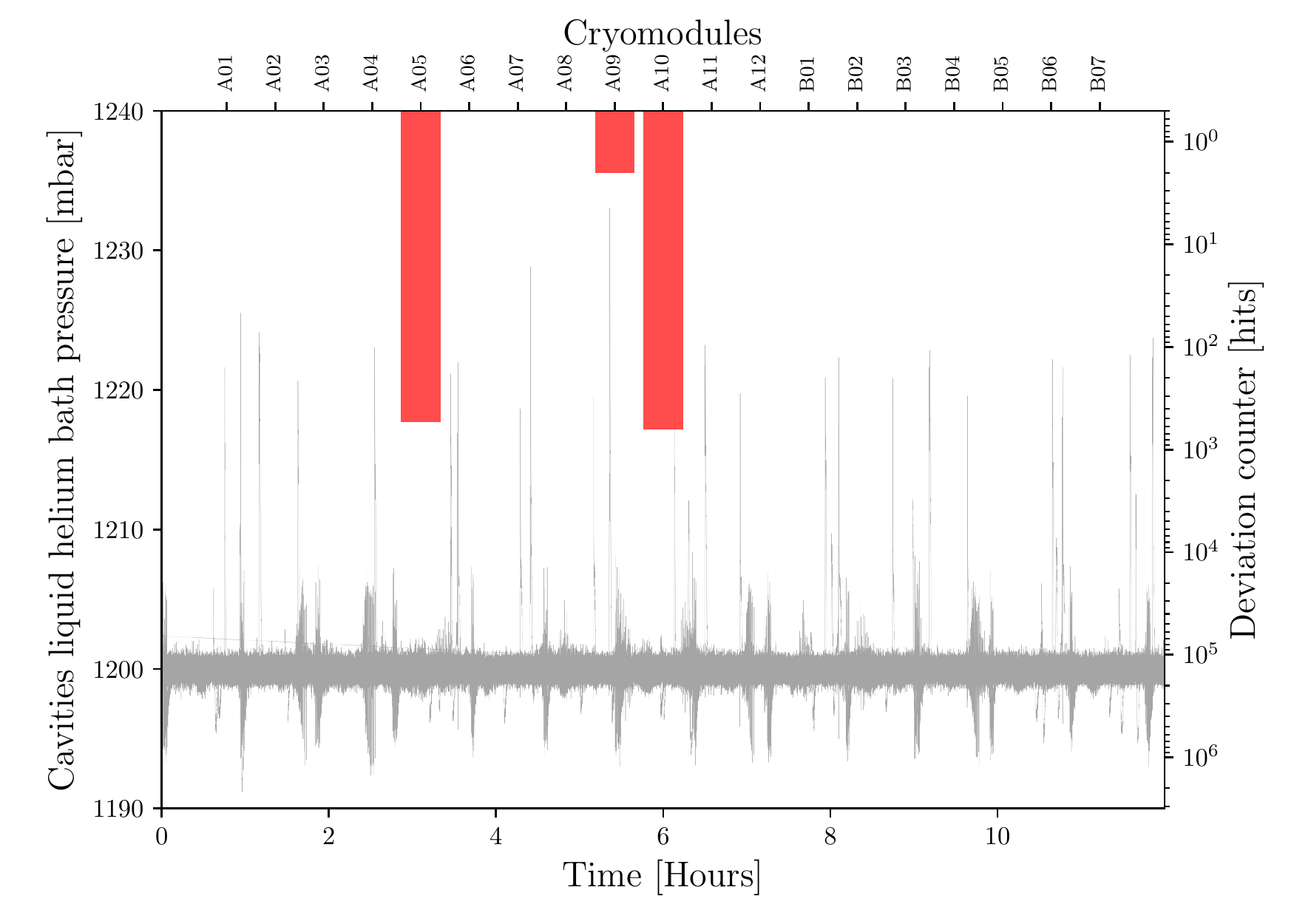}
  \caption{Pressure fluctuations (gray) and $\pm~5~mbar$ deviation counter (red). Data of :  2017-11-27 | start 20:00.}\label{ptc_1127}
\end{figure}

\begin{figure}[hbt]
  \includegraphics[width=.5\textwidth, right]{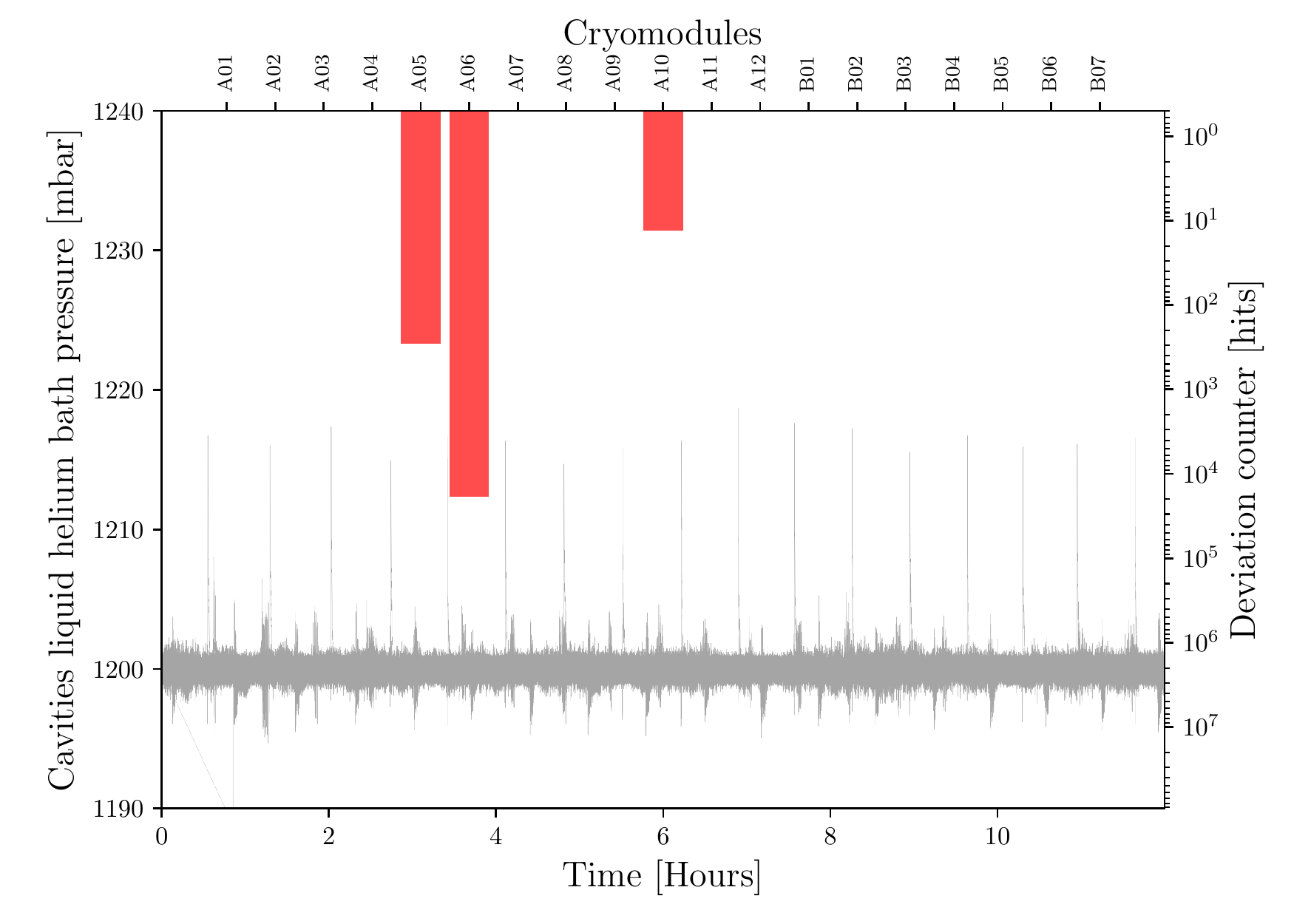}
  \caption{Pressure fluctuations (gray) and $\pm~5~mbar$ deviation counter (red). Data of :  2017-11-28 | start 20:00.}\label{ptc_1128}
\end{figure}

\begin{figure}[hbt]
  \includegraphics[width=.5\textwidth, right]{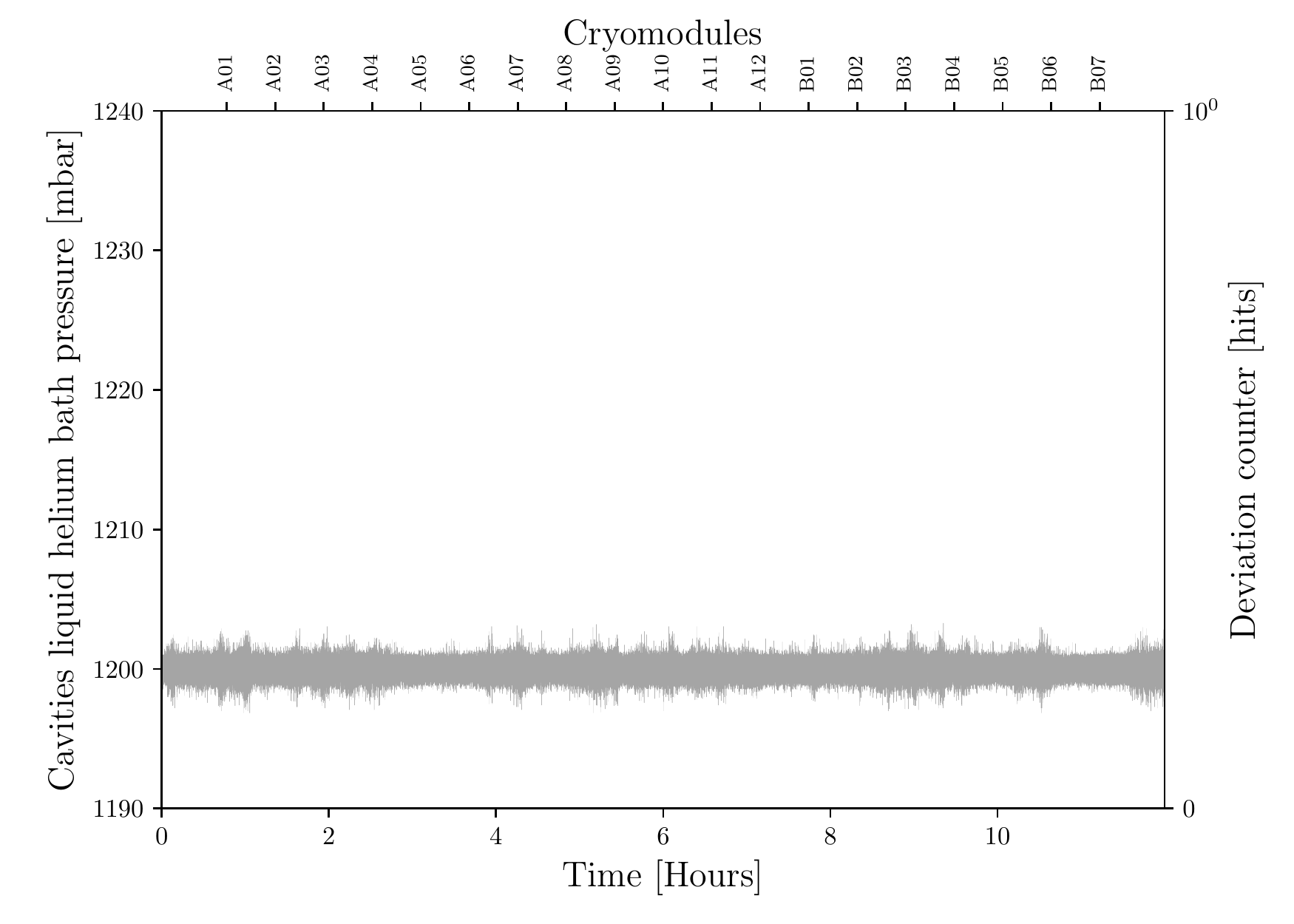}
  \caption{Pressure fluctuations (gray) and $\pm~5~mbar$ deviation counter (red but none measured). Data of :  2017-12-05 | start 20:00.}\label{ptc_1205}
\end{figure}

We then monitored every night the pressure fluctuations and counted the number of deviations from the accepted pressure range ($\pm~5~mbar$). While stabilised most of the time, the pressure showed periodic or pseudo-periodic instabilities that seemed to move from one LINAC position to another. This behaviour seemed to be more frequent for the 12 first type A cryomodules. Figure \ref{ptc_1127} shows such behaviour for CMA05 and CMA10 and figure \ref{ptc_1128} for CMA06. Other periods were more stable and did not show any instability over $\pm~5~mbar$ (see figure \ref{ptc_1205} for example).

\begin{figure}[hbt]
  \includegraphics[width=.5\textwidth, right]{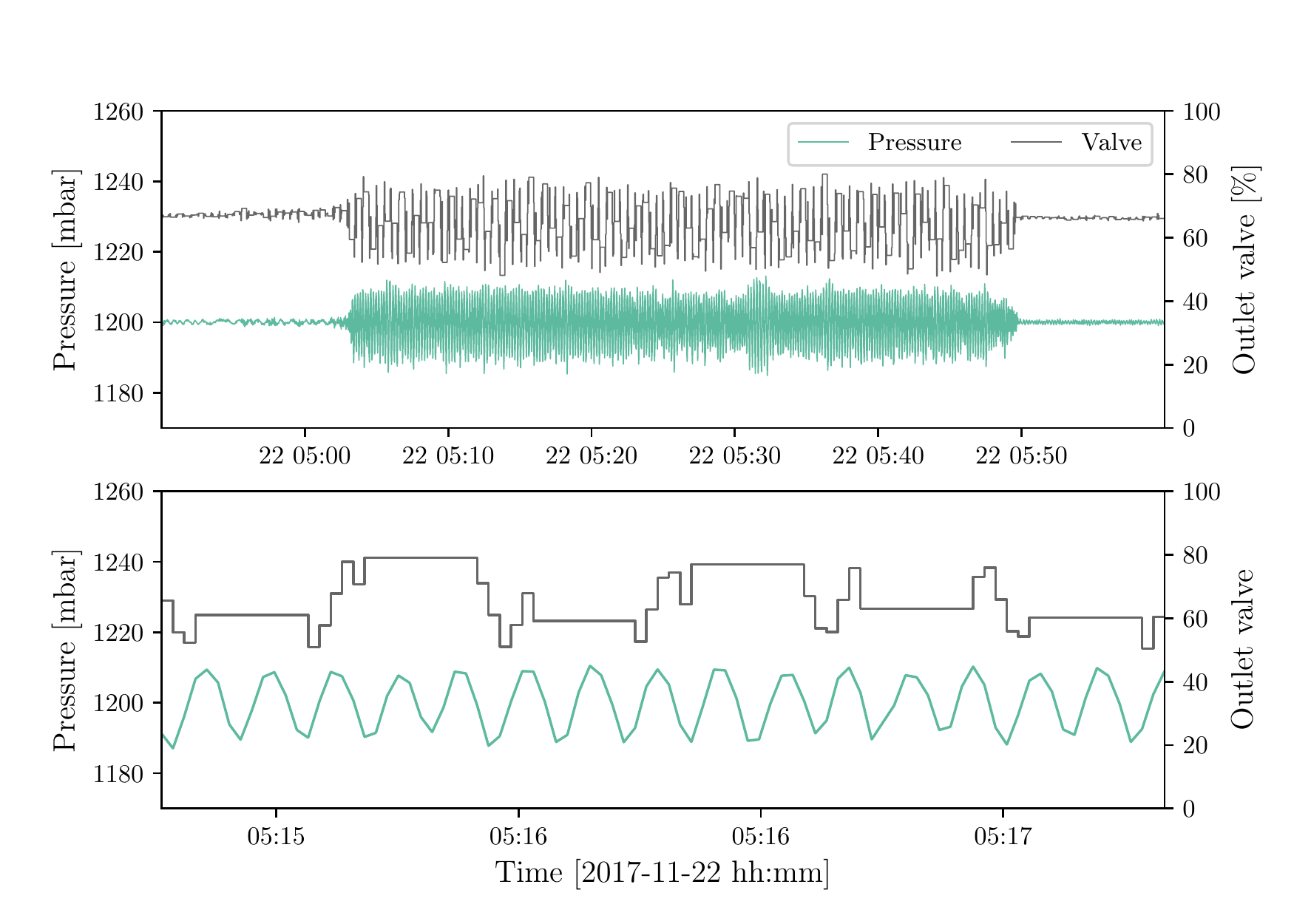}
  \caption{2017-11-22 | CMA01}\label{pt_fv_cma01}
\end{figure}

When looking in more details for the action of the pressure control valve during instabilities, we could see that instabilities seemed uncorrelated to the actions of the valves (see figure \ref{pt_fv_cma01}). Moreover, the maximum sampling rate of pressure data acquisition (300 ms) couldn't allow us to distinguish any reproducible behaviour with a fixed periodicity. All this pointed to the presence of hidden fast phenomena distributed all over the LINAC.

\subsection{Other measurements}
\begin{figure}[hbt]
  \includegraphics[width=.5\textwidth, right]{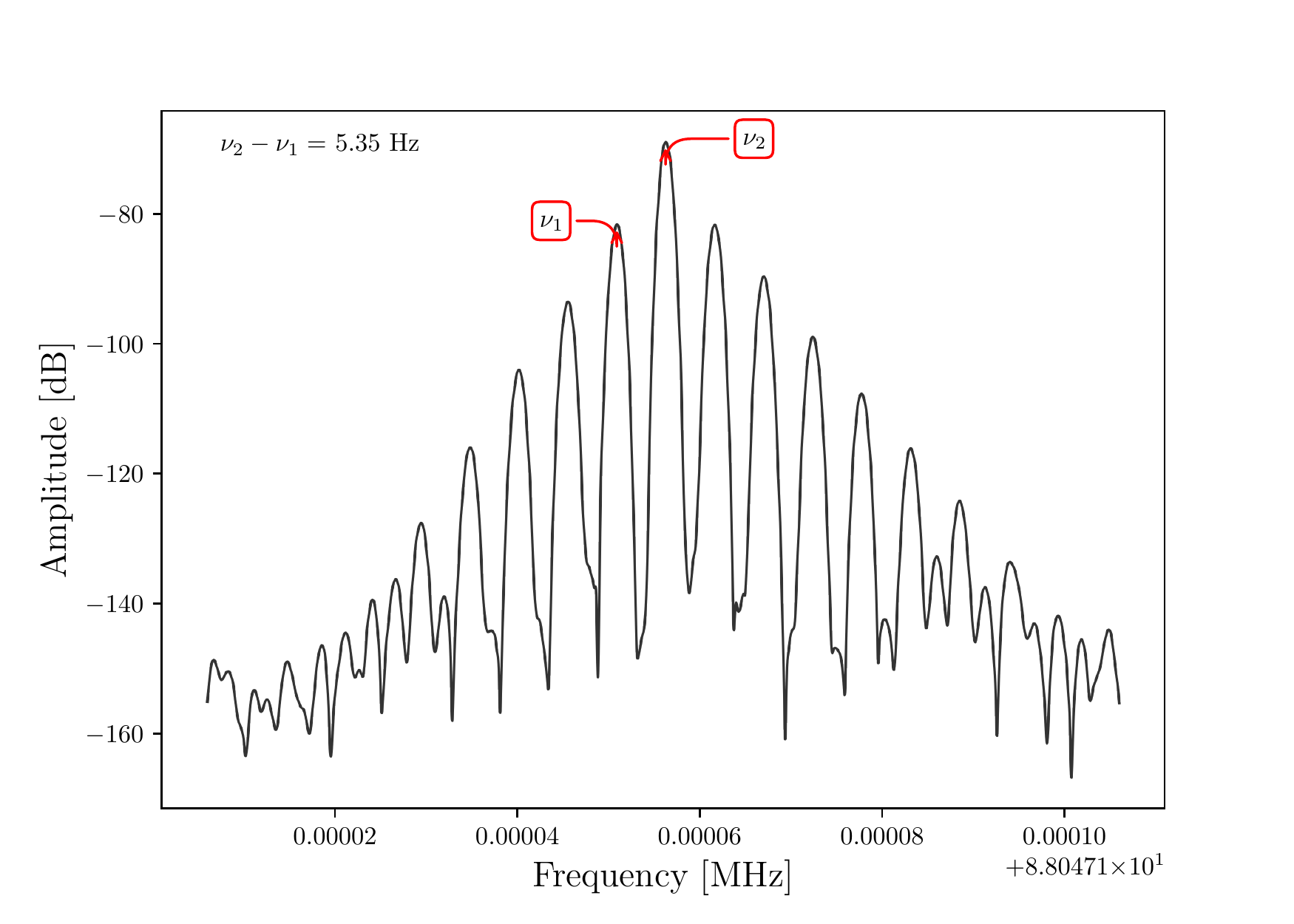}
  \caption{Transmitted power modulation measurement of cavity number 25 (CMB07 cavity 1). Cryomodule was cold, thermalised and regulated both in liquid helium level and pressure. Pressure of liquid helium bath, as measured, was stable within $\pm~5~mbar$. Modulation frequency was measured to be $5.35~Hz$}\label{cmb07_rf}
\end{figure}

During this 2017 run, first RF measurements of the cavities have been performed.  This was a first cold measurement of some RF properties of the cavities in the LINAC. Due to the distance between the LINAC and the generators (power loss) and the limitations on the injected RF power in the LINAC (nuclear safety authorisation was pending), we used an external RF synthesizer to measure and characterise the transmitted power response. First, a self-oscillating closed loop circuit was used to find the resonance of the cavities. Then a low power frequency generator was used to inject the RF signal and a Yokogawa network analyser was used to analyse the output. Surprisingly, both an amplitude and a frequency modulation was noticed. These measurements have been performed on CMA07, CMA11 and CMB07. Figure \ref{cmb07_rf} shows an amplitude modulation spectrum with peak to peak distance of 5.35 Hz. Other measurements consistently showed a modulation between 4 and 6 Hz. Frequencies of modulation seemed specific to the measured cryomodules as repeated measurements on the same cryomodules showed the exact same modulation frequencies. These modulations were not noticed during vertical cryostat measurements. This might mean that these fluctuations could be due to the final setup of the cryomodules in the LINAC. In a final beam configuration, the modulations would be compensated by the amplifiers power margin. However, the amplitude of these modulations are high enough to cause the system to hit its limits. Therefore, the origin of these microphonics will have to be investigated further and their effect mitigated.

The measured frequencies were too low to be a consequence of the RF test bench and seemed to be more consistent with vibrations of a mechanical nature. Several accelerometer measurements of vibrations have then been performed in the LINAC at different locations and nothing unusual has been noticed. Although few Hz seemed usually low also to be acoustic like oscillations, this could explain the fast fluctuations of the pressures in the LINAC.

\section{Conclusion}

After countless developments and installation efforts, the SPIRAL 2 superconducting LINAC was finally cooled down for the first time in October 2017. It was also the first time that all cryogenic systems were tested in a configuration as close as possible to the operation configuration. Cooling down such a system with diphasic helium and with no by-pass at the end of the LINAC proved to be a challenge. However, trials showed that it is possible to find an optimised cool down strategy with minimal effects on the performances of the cavities. Pressure and level regulations with simple yet optimised PI controllers were successful for time laps as long as 12 hours. Pseudo-periodic instabilities on all cavities as well as possible hidden fast fluctuations will have to be investigated further. Another limitation of this run was that all cavities were characterised in a so called static configuration. It means that no real nor simulated RF heat load could be taken into account. Low power RF measurements of the cavities showed resonances at their design values. However,  amplitude and frequency modulations of the transmitted power were detected and measured. They indicated important microphonics that could be linked to the cryogenic system and explain some of the unexplained pressure fluctuations that have been observed. Investigations of the measured microphonics were continued with a dedicated bench that allows the synchronous measurements of RF and pressure modulations. This cool down was also important to gather precious data for precise thermodynamic modelling of the SPIRAL 2 LINAC\cite{Vassal_2019}. The next cool down will hopefully benefit of a model based pressure and liquid helium level control as well as optimised input cool down parameters. It will also benefit of a dedicated bench for synchronous measurements of RF and piezoelectrically measured pressure fluctuations. 

\bibliography{ref}

\end{document}